\documentclass[journal=jacsat,manuscript=article]{achemso}
\pdfoutput=1

\usepackage[version=3]{mhchem}

\usepackage{amsmath}
\usepackage{braket}

\usepackage{subfig}
\usepackage{caption}
\usepackage{wrapfig}
\usepackage{multirow}
\usepackage{tikz}

\author{Kuntal Talit}
\affiliation{Department of Physics, University of California, 
Merced, 5200 N. Lake Rd., Merced, CA 95343}

\author{David A. Strubbe}
\email{dstrubbe@ucmerced.edu}
\phone{+1-209-228-4481}
\affiliation{Department of Physics, University of California, 
Merced, 5200 N. Lake Rd., Merced, CA 95343}
\SectionNumbersOn
\date{\today}

\title[]
  {Quantifying Hidden Symmetry in the Tetragonal CH$_3$NH$_3$PbI$_3$ 
  Perovskite}

\begin{document}
%%%%%%%%%%%%%%%%%%%%%%%%%%%%%%%%%%%%%%%%%%%%%%%%%%%%%%%%%%%%%%%%%%%%%
%% TOC graphic
%%%%%%%%%%%%%%%%%%%%%%%%%%%%%%%%%%%%%%%%%%%%%%%%%%%%%%%%%%%%%%%%%%%%%
\begin{tocentry}
\label{For Table of Contents Only}
    \includegraphics[width=5cm]{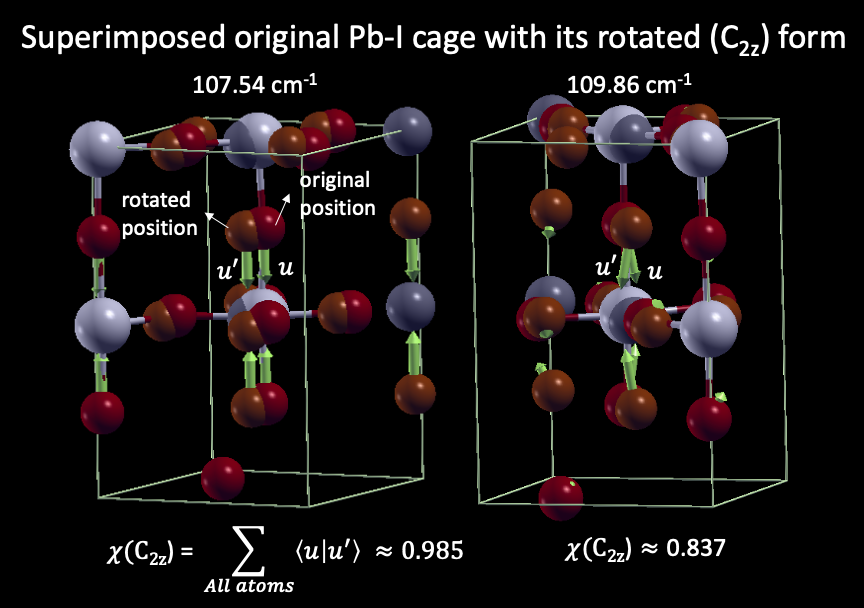}
\end{tocentry}
%%%%%%%%%%%%%%%%%%%%%%%%%%%%%%%%%%%%%%%%%%%%%%%%%%%%%%%%%%%%%%%%%%%%%
%% Abstract
%%%%%%%%%%%%%%%%%%%%%%%%%%%%%%%%%%%%%%%%%%%%%%%%%%%%%%%%%%%%%%%%%%%%%
\begin{abstract}
 The assignment of an exact space group to the tetragonal CH$_3$NH$_3$PbI$_3$ perovskite structure is experimentally challenging and controversial in the literature. Average orientation of the methylammonium ion that gives symmetry to the experimental measurement is not captured in a static density functional theory calculation, although the quasi-I4cm and quasi-I4/mcm structures are commonly used in calculations. In this work we have developed a methodology to quantify the hidden symmetry of these structures using group theory, to enable use of symmetries in understanding spectroscopy and other properties. We study the approximate symmetry of vibrational modes, including analysis of degenerate representations, as well as the dielectric, elastic, electro-optic, Born effective charge, and Raman tensors and the dynamical matrix. Comparing to each subgroup of the full tetragonal $D_{4h}$, our results show that the quasi-I4cm is best described by the expected corresponding point group C$_{4v}$, whereas the quasi-I4/mcm (despite corresponding to point group $D_{4h}$) is best described by the lower symmetry C$_{2v}$. Our methodology can be useful generally for analysis of other soft hybrid materials or any approximately symmetric material.
\end{abstract}

%%%%%%%%%%%%%%%%%%%%%%%%%%%%%%%%%%%%%%%%%%%%%%%%%%%%%%%%%%%%%%%%%%%%%
%% Main part of the manuscript
%%%%%%%%%%%%%%%%%%%%%%%%%%%%%%%%%%%%%%%%%%%%%%%%%%%%%%%%%%%%%%%%%%%%%

\begin{figure}[ht]
    \centering
    \includegraphics[width=\textwidth]{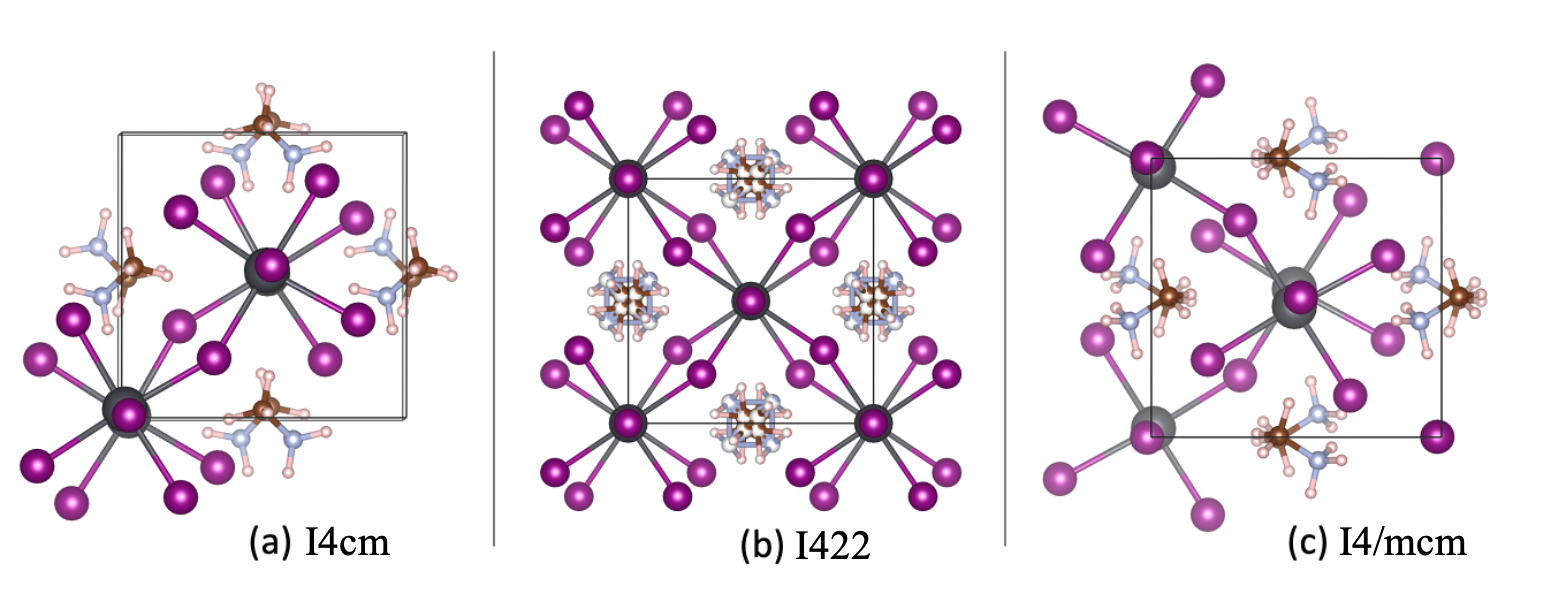}
    \caption{Tetragonal MAPI with different space group symmetries: (a) I4cm ($C_{4v}$) structure, (b) experimental structure ($D_{4}$) with partial occupancies, and (c) I4/mcm ($D_{4h}$), having the full symmetry of the tetragonal structure.}
    \label{fig:intropic}
\end{figure}

Hybrid organometallic perovskites are one of the most researched material for solar cell application in last decade. There are more than sixteen thousand research documents published between 2009 to 2019 regarding perovskite solar cell.\cite{shikoh2020quantitative} A huge amount of research has been done towards low-cost fabrication, increasing the photo-conversion efficiency, making active layer materials etc. But in-depth understanding about some fundamental aspects is still missing. The exact symmetry of room-temperature tetragonal methylamonium lead iodide (MAPI) is one of them. There is still debate about the space group symmetry of the tetragonal MAPI. Some reports suggest that the structure is ferroelectric or polar having quasi-I4cm space group symmetry\cite{stoumpos2013semiconducting,xie2015study} while others have found tetragonal MAPI structures that are antiferroelectric or antipolar in nature, having quasi-I4/mcm space group symmetry\cite{poglitsch1987dynamic,PND,franz2016interaction}(Fig. \ref{fig:intropic}) There are also reports that identified the space group as I4/m\cite{baikie2013synthesis}. There are experiments that reports the structure to have space groups I422 and P$4_22_12$ which are subgroups of I4/mcm.\cite{arakcheeva2016ch3nh3pbi3}\par 
It is important to know the symmetry of the tetragonal structure because symmetry is an essential tool to understand the spectroscopy and other properties of hybrid perovskites.\cite{pedestrian} An important example relates to the electronic properties: the two different structures I4cm and I4/mcm have different electronic properties. I4/mcm is a centrosymmetric structure with inversion symmetry and theoretically it should not produce Rashba splitting in the bandstructure while I4cm is a non-centrosymmetric structure without the inversion symmetry and it produces significant Rashba splitting.\cite{Frohna2018} A DFT study concluded that any calculated significant Rashba splitting in case of the I4/mcm structure is incorrect and may be due to incorrect structural relaxation.\cite{Frohna2018} Another DFT study determined that the energy difference between a quasi-I4cm and a quasi-I4/mcm structure is very small (0.1 eV) and they can coexist in a single crystal with domains of altering tilting directions which can further dynamically interchange into each other at room temperature crossing some energy barrier that caused due to the specific interaction between the MA$^+$ ion and the inorganic cage\cite{quarti2014interplay}. A source of difficulty in determining the symmetry experimentally is that when the material goes through a phase transition from cubic to tetragonal structure due to temperature changes, it loses some symmetry elements which can gives rise to twinning along the lost symmetry element.\cite{breternitz2020twinning} In experiment, we mainly get the overall symmetry of the whole crystal, but sometimes the unit cell might have different symmetry due to such twinning within the crystal.\par
In case of the experimental structure, each MA$^+$ ion is statistically distributed with a fractional occupation of 25\% for each of 4 orientation.\cite{arakcheeva2016ch3nh3pbi3} This arrangement provides symmetry. To do any theoretical calculation we must take a snapshot of multiple possible orientations of the MA$^+$ ion and at that moment we lose all the symmetry.  Quarti \textit{et al.} have done a detailed study of the tetragonal structure and found that a set of polar structures (I4cm) are more stable than the apolar (I4/mcm) ones.\cite{quarti2014interplay} The most stable structure reported in their study is a polar structure with I4cm space group symmetry and this structure is used by other works\cite{brivio2015lattice,fan2015ferroelectricity} (though seems to be described as I4/mcm).\par

In case of the low-temperature orthorhombic structure, 4 MA$^+$ ions in the unit cell are static which gives it a perfect D$_{2h}$ symmetry. At high temperatures, the random spinning of the MA$^+$ ion within the cage makes the structure pseudo-cubic, and is not even close to any symmetry, complicating theoretical analyses.\cite{talit} For the room-temperature tetragonal structure, the average over space and time of this random spinning makes this structure quasi-I4cm or quasi-I4/mcm. So, the tetragonal MAPI does not have any exact symmetry, but is considered to have approximate symmetry. Previous literature however has not quantitatively assessed the symmetry of these structures, to describe rigorously how close they are to I4cm, I4/mcm, or any other space group. In this work, We want to quantify how well symmetries such as I4cm or I4/mcm describe the structures, and find the highest degree of approximate symmetry that can be used to describe properties of this tetragonal structure.  To identify the hidden symmetry in the structure we have checked the symmetry from different aspects: (a) atomic coordinates, (b) vibrational modes, (c) elastic tensor (or stiffness matrix), (d) dielectric tensor, (e) electro-optic tensor, (f) dynamical matrix, (g) Born effective charge tensors, and (h) atomic Raman tensors. The elastic, dielectric, and electro-optic tensors provide global mechanical and electronic properties, whereas the coordinates and dynamical matrix provide atom-resolved structural properties, and the Born effective charges and atomic Raman tensors provide atom-resolved mixed structural/electronic properties. We use these assessments of symmetry in structural, electronic, and vibrational aspects in combination to identify the most appropriate symmetry subgroup description of the tetragonal structure.\par

For any crystal structure that has exact symmetry vibrational modes can be classified according to irreducible representations, but this cannot be done when the structure does not have any symmetry. In this work, we have developed a method to calculate the approximate irreducible representation of the vibrational modes for an approximately symmetric structure. We use the approximate symmetry of the crystal structure and its character table as our input and use group theory to calculate approximate characters in the character table and thereby calculate the irreducible representations of the vibrational modes. We have calculated the contributions of irreducible representations for each phonon mode of tetragonal MAPI which can be helpful for spectroscopic studies. As a test of our methodology, we have also calculated the same for perfectly symmetric orthorhombic MAPI and TiO$_2$ and it gives correct irreducible representations for both the systems compared with Quantum ESPRESSO results. Our methodology can be useful to calculate hidden symmetry and approximate mode irreducible representations for any approximately symmetric structure.\par

%%%%%%%%%%%%%%%%%%%%%%%%%%%%%%%%%%%%%%%%%%%%%%%%%%%%%%%%%%%%%%%%%%%%%
%%\section{Main Body}
%%%%%%%%%%%%%%%%%%%%%%%%%%%%%%%%%%%%%%%%%%%%%%%%%%%%%%%%%%%%%%%%%%%%%
We have studied two different tetragonal structures, quasi-I4cm\cite{brivio2015lattice} and quasi-I4/mcm\cite{leppert2016electric}, using density functional theory. Computational details, similar to our previous work on strain effects in cubic MAPI,\cite{talit} are given in the supporting information.
%%%%%%%%%%%%%%%%%%%%%%%%%%%%%%%%%%%%%%%%%%%%%%%%%%%%%%%%%%%%%%%%%%%%%
%%\subsection{Symmetry in the crystal structures}
%%%%%%%%%%%%%%%%%%%%%%%%%%%%%%%%%%%%%%%%%%%%%%%%%%%%%%%%%%%%%%%%%%%%%
To check the initial symmetry of the structures we have used FINDSYM\cite{stokes2017findsym,stokes2005findsym}. The result is given in Table S1. As we already know that the theoretical structure does not have any symmetry due to the different orientations of the MA$^+$ ions within the structure, we have removed all the MA$^+$ ions from the I4cm structure and checked the symmetry of the Pb-I cage only. With some tolerance with respect to the lattice and the atomic positions, we found that the Pb-I cage still holds the D$_{4h}$ point group symmetry. One thing to notice here is that the Pb-I cage and the whole structure both have symmetry $C_s$ which is a subgroup of D$_{4h}$, even with low tolerance values. We will come back to this point while explaining phonon mode symmetries. For the quasi-I4/mcm structure, even with low tolerance values, the predicted symmetry by FINDSYM is C$_{2v}$ which is orthorhombic symmetry and lower in symmetry than D$_{4h}$. This gives an indication that the tetragonal structure may be better described using some lower symmetry subgroups of D$_{4h}$.\par

%%%%%%%%%%%%%%%%%%%%%%%%%%%%%%%%%%%%%%%%%%%%%%%%%%%%%%%%%%%%%%%%%%%%%%%
%%\subsection{Symmetry in elastic tensors}
%%%%%%%%%%%%%%%%%%%%%%%%%%%%%%%%%%%%%%%%%%%%%%%%%%%%%%%%%%%%%%%%%%%%%%%
Next, we consider three tensors which provide global (not atom-resolved) properties of the system, the elastic, dielectric, and electro-optic tensors. Examining the full stiffness tensor $C_{ij}$, $6 \times 6$ in Voigt notation, shows symmetry in mechanical response. Our calculated tensors for quasi-I4cm and the quasi-I4/mcm structures are shown in Fig. S1. For tetragonal (I) crystal system\cite{mouhat2014necessary} we should have nonzero elements $C_{11}= C_{22}$, $C_{33}$, $C_{44}= C_{55}$, $C_{66}$, $C_{12}$, and $C_{13}= C_{23}$. The stiffness tensor for quasi-I4cm structure closely follows the tetragonal (I) symmetry, except there are small off-diagonal values. For the stiffness tensor of the quasi-I4/mcm structure, all the diagonal values are different and $C_{13}$ is not same as $C_{23}$. This is not even close to tetragonal (I) symmetry, but more like orthorhombic symmetry. Applying symmetry rotations that belongs to D$_{4h}$ point group to the stiffness matrix it is possible to quantify how each symmetry is obeyed by the stiffness matrix of both the structures.\par
 
%%%%%%%%%%%%%%%%%%%%%%%%%%%%%%%%%%%%%%%%%%%%%%%%%%%%%%%%%%%%%%%%%%%%%%%
%%\subsection{Symmetry in dielectric tensors}
%%%%%%%%%%%%%%%%%%%%%%%%%%%%%%%%%%%%%%%%%%%%%%%%%%%%%%%%%%%%%%%%%%%%%%%
We have calculated the static electronic ($\epsilon_\infty)$ and electronic+ionic contribution ($\epsilon_0$) of the dielectric tensor for our tetragonal MAPI structures (Fig. S2). Both show similar symmetry properties. For a perfectly symmetric tetragonal structure we should have only the diagonal values with ($\epsilon_{11}=\epsilon_{22}$). Calculated off-diagonal values are also an indication that the structure is not properly symmetric. Although the off-diagonal elements are close to zero for I4/mcm structure, I4cm obeys the tetragonal symmetry better than I4/mcm. The dielectric tensor for the I4cm structure also is consistent with $S_4$, $D_{2d}$, $C_4$, $C_{4v}$, $D_4$ and $D_{4h}$ point group symmetries. For I4/mcm, the dielectric tensor is consistent with $C_{2v}$, $D_2$, and $D_{2h}$ point group symmetries.\par

%%%%%%%%%%%%%%%%%%%%%%%%%%%%%%%%%%%%%%%%%%%%%%%%%%%%%%%%%%%%%%%%%%%%%%%
%%\subsection{Symmetry in electro-optic tensor}
%%%%%%%%%%%%%%%%%%%%%%%%%%%%%%%%%%%%%%%%%%%%%%%%%%%%%%%%%%%%%%%%%%%%%%%
The static non-linear electro-optic tensor $\chi^{(2)}$ is a sensitive probe of symmetry, particularly centrosymmetry,\cite{Frohna2018} since all tensor elements would vanish in the presence of exact inversion symmetry. The calculated values are in Rydberg atomic units. For the quasi-I4/mcm structure all the values are close to zero except for $\chi^{(2)}_{zzz}$, $\approx 30.19 \ {\rm a.u.}$. More values are nonzero for I4cm, a clear indication that it is non-centrosymmetric. We have further checked all the symmetries that $\chi^{(2)}$ for a non-centrosymmetric structure should obey\cite{shen1984principles}. We see that $\chi^{(2)}_{zzz}=22.73\ {\rm a.u.}$ is the largest element, and $\chi^{(2)}_{xxx} \approx \chi^{(2)}_{yyy}\approx 3.125 \ {\rm a.u.}$ are other large elements, which are supposed to be zero for all tetragonal symmetries. These nonzero values are consistent with $C_s$, $C_4$ and $C_{4v}$, and $D_{4h}$ point groups.\par

%%%%%%%%%%%%%%%%%%%%%%%%%%%%%%%%%%%%%%%%%%%%%%%%%%%%%%%%%%%%%%%%%%%%%%%
%%\subsection{Symmetry in Born Effective Charges, 
%%  Raman tensors and dynamical matrices}
%%%%%%%%%%%%%%%%%%%%%%%%%%%%%%%%%%%%%%%%%%%%%%%%%%%%%%%%%%%%%%%%%%%%%%%
\begin{figure}[ht!]
    \centering
    \includegraphics[width=0.99\textwidth]{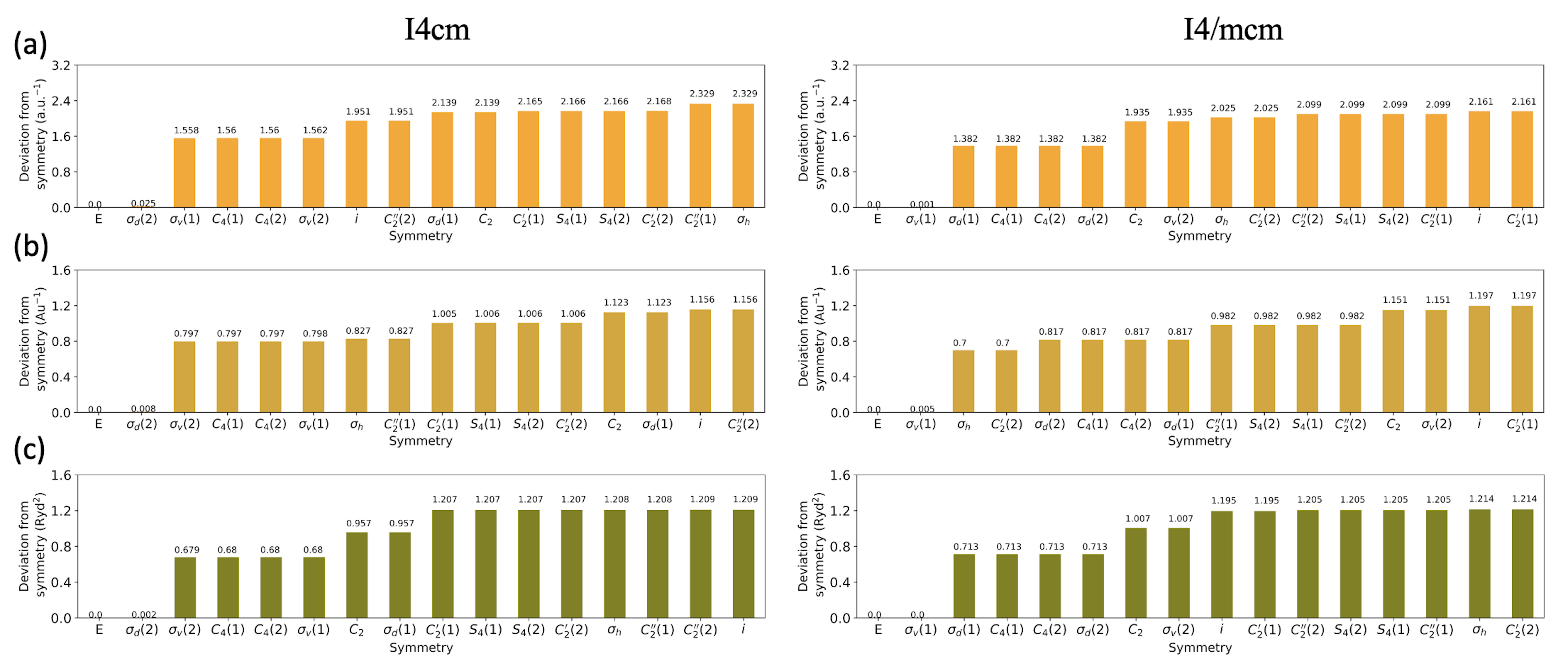}
    \caption{Deviation from symmetry for each operation of $D_{4h}$ for (a) Born effective charges, $\Delta_Z$, (b) atomic Raman tensors, $\Delta_R$, and (c) dynamical matrices, $\Delta D$, of I4cm and I4/mcm structures.}
    \label{fig:devsym}
\end{figure}

%%%%%%%%%%%%%%%%%%%%%%%%%%%%%%%%%%%%%%%%%%%%%%%%%%%%%%%%%%%%%%%%%%%%%%%
%%\subsection{Symmetry in vibrational modes}
%%%%%%%%%%%%%%%%%%%%%%%%%%%%%%%%%%%%%%%%%%%%%%%%%%%%%%%%%%%%%%%%%%%%%%%
Our tetragonal structures in this work do not have any exact symmetry and hence no irreducible representations for their vibrational modes. We can still calculate the approximate mode irreducible representations using help of group theory. The well known formula for decomposition of reducible representation into its corresponding irreducible representations is given in Eq. \ref{eq:decomp_rrp}\cite{harris1989symmetry}. The number of times the irreducible representation $\Gamma_j$ appears in the reducible representation is given by $a_j$, where $h$ is the order of the point group, $C_k$  denotes a class in the point group, $N_k$ is the number of elements in $C_k$ and $\chi^{(\Gamma_j)}(C_k)$ represents the character of the irreducible representation $\Gamma_j$ for a symmetry operation in class $C_k$.
\begin{equation}\label{eq:decomp_rrp}
    a_j=\frac{1}{h}\sum_k N_k[\chi^{(\Gamma_j)}(C_k)]^* \chi (C_k)
\end{equation}
The second orthogonality rule for the columns of the character table is given in Eq. \ref{eq:2nd_orthogonality}.

\begin{equation}\label{eq:2nd_orthogonality}
     \sum_{j} [\chi^{(\Gamma_j)}(C_k)]^* \chi^{(\Gamma_j)} (C_{k'})=\frac{h}{N_k} \delta_{kk'}
\end{equation}
For $k=E$ (identity operation), $N_k=1$. So we can rewrite Eq. \ref{eq:2nd_orthogonality} as 
\begin{equation}
    \sum_{j} [\chi^{(\Gamma_j)}(E)]^* \chi^{(\Gamma_j)} (C_{k'})=h \delta_{Ek'}
\end{equation}

$\chi^{(\Gamma_j)}(E)=1$ for A or B (non degenerate) irreducible representation, $\chi^{(\Gamma_j)}(E)=2$ for E (doubly degenerate) and $\chi^{(\Gamma_j)}(E)=3$ for T (triply degenerate) irreducible representations.

Multiplying Eq. (\ref{eq:decomp_rrp}) with $\chi^{(\Gamma_j)}(E)$ and summing over $j$ we get
\begin{equation}\label{eq:sum_aj_explain}
\begin{aligned}
        \sum_{j}\chi^{(\Gamma_j)}(E) a_j &=\frac{1}{h}\sum_k\sum_{j} \chi^{(\Gamma_j)}(E)[\chi^{(\Gamma_j)}(C_k)]^* \chi (C_k)\\
  & = \frac{1}{h}\sum_k h \delta_{Ek'} \chi (C_k)\\
  & =\chi (E)
\end{aligned}
\end{equation}

It is interesting to note that when we sum over the contributions ($a_j$) of all irreducible representations for any mode, it turns out exactly 1 for non-degenerate, 2 for doubly degenerate, and 3 for triply degenerate modes. To make the sum 1 for all the modes we have to divide $\chi(E)$ by 2 for for doubly degenerate, and by 3 for triply degenerate modes.We have used Eq. \ref{eq:decomp_rrp} to calculate $a_j$ and then use Eq. \ref{eq:sum_aj_explain} to find out the proportion of irreducible representations for each mode.\par

\begin{figure}[ht!]
    \centering
    \includegraphics[width=0.7\textwidth]{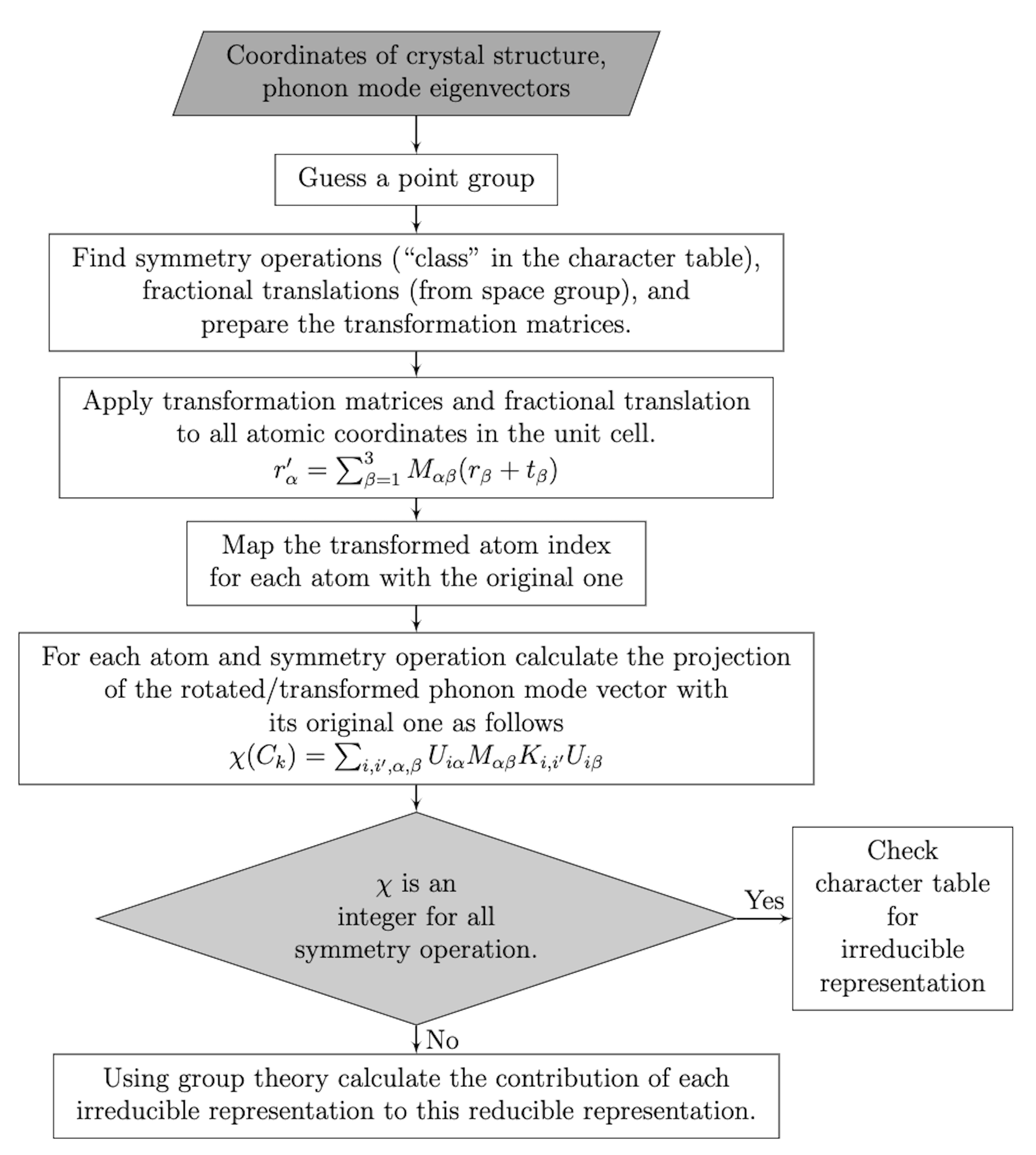}
    \caption{Approximate phonon mode symmetry calculation flow chart.}
    \label{Fig:FC}
\end{figure}
 
Using the above methodology, we have calculated the approximate irreducible representations for 
I4cm and I4/mcm structures. The main process is explained in the form of a simple flow chart (Fig. \ref{Fig:FC}). We started with the I4cm structure. We have relaxed the structure using as mentioned in the computational method section. Density functional perturbation theory (DFPT) is used to calculate the phonon modes at $q=0$. The acoustic sum rule (ASR) is applied using the dynmat.x code as implemented in Quantum ESPRESSO. We have taken the position coordinates of the relaxed structure and its calculated phonon mode vectors as input. The closest symmetry of the structure we considered is I4/mcm (or D$_{4h}$ point group), because this is the highest symmetry in tetragonal structure and if we calculate this once, we can always get results for I4cm as it is a subgroup of I4/mcm. From the character table of $D_{4h}$, we get all the symmetry operations (16 in our case) and the target irreducible representations\cite{gelessus1995multipoles}. From the space group we find all the fractional translations that are involved. We have constructed all the $3\times3$ rotational matrices ($M_{\alpha\beta}$) and the fractional translation vector ($\vec{t}$) to apply on the original atomic positions ($\vec{r}$) of the crystal unit cell as $r'_{\alpha}=\sum_{\beta=1}^3 M_{\alpha\beta}(r_\beta +t_\beta)$ where $\alpha$ and $\beta$ denote the $x,y,$ and $z$ directions. To make the calculations simple, we started with the Pb-I cage only. The orientation of the MA$^+$ ions in the structure breaks the symmetry, but the Pb-I cage still holds the D$_{4h}$ point group symmetry within certain tolerance values (Table S1).
After removing the MA$^+$ ion from the structure, we have applied all the symmetry operations on the structure to find how they swap atoms. When we apply a rotation to the crystal structure, if for example, a carbon atom (C1) takes the place of another carbon atom (C2), we say C1 and C2 are swapped atoms of each other with respect to that rotation. Vibration modes should obey certain symmetry operations based on the symmetry of the crystal structure. We apply the symmetry transformation to the vibrational mode Cartesian vectors and calculate the projection of the transformed mode vector to the original one for each atom and the value will give us the character value $\chi$ corresponding to that symmetry class for that mode. The equation for calculating the projection is 
\begin{equation}\label{eq:Chi}
    \chi(C_k)=\sum_{i,i',\alpha,\beta} U_{i\alpha} M_{\alpha\beta}(C_k) K_{i,i'}(C_k) U_{i\beta}
\end{equation}
where $i$ and $i'$ denote the atom index of the original and transformed atoms respectively, $K_{i,i'}(C_k)$ denotes the matrix that transforms $i$ to $i'$, $U_{i\alpha}$ denotes the mode vector for atom $i$ in direction $\alpha$, and $C_k$ denotes the symmetry class for which $\chi$ is been calculated.\par

To calculate the character we need the mode eigenvector for that particular mode. Once we remove the MA$^+$ ions we need to re-normalize the mode vectors ($\vec{U_i}$) for Pb and I as $\vec{V_i}=\vec{U_i} /\sqrt{\sum_{i=1}^{N}\lvert{\vec{U_i}\rvert^2}}$ where $i$ is the atom index and $N$ is the total number of atoms after removing all the MA$^+$ ions. We calculated the value of $\chi$ for all symmetry classes and for each mode of the tetragonal MAPI. As our structure is not exactly symmetric, we did not expect to get integer values for $\chi$ for all the symmetry classes, in fact our calculated values are in fractions. So, we need to find a different way rather than checking character table for a direct match as we have already mentioned in the flowchart(Fig.\ref{Fig:FC}).\par

For each phonon mode we have calculated $\chi(C_k)$ for all symmetry class $C_k$ belonging to the point group $D_{4h}$ and prepared a table which we call calculated character of modes because it is like a character table but with character values in fractions rather than in integers as we normally see in a character table. Each row of this calculated character of mode table is treated as a reducible representation and we decompose them into the irreducible representations using group theory (Eq. \ref{eq:decomp_rrp}).\par

We noticed that the sum of contributions of all the irreducible representations become 1 for all the modes. We gave a theoretical explanation why this occurs using group theory (Eq. \ref{eq:sum_aj_explain}). If we just sum up the contributions ($a_j$) for all irreducible representations we end up getting sum as 2's and 3's for doubly degenerate or triply degenerate modes which is a problem because in that case the contributions of irreducible representations for each mode do not sum up to one, which makes it hard to compare between all the modes. We have studied it further by decomposing the degenerate modes into a possible combination of two symmetrized non-degenerate modes by looking at how the basis functions ($x$,$y$) transform with different symmetry operations and  repeated the same calculations for calculating $a_j$ and this time it gave the sum as 1 but our symmetrized combination is just one of the many possible permutations of how ($x$,$y$) basis can transform under the symmetry operations. It become even harder when the character value become imaginary in some cases, for example, in the character table for C$_3$ point group the degenerate irreducible representation is a symmetrized combination of 1, $e^{i2\pi/3}$, and $e^{-i2\pi/3}$. This issue is known as the doubling problem\cite{carter1993representations}. On the other hand, we see that in our formulation of Eq. \ref{eq:sum_aj_explain} we just have to divide the sum by 2 for the doubly degenerate mode as  $\chi^{(\Gamma_j)}(C_E)=2$ and this makes the sum of all the irreducible representations for each mode (including the degenerate ones) as 1. We use this treatment, in which case we do not need to split the degenerate mode into an arbitrary basis.\par

%%%%%%%%%%%%%%%%%%%%%%%%%%%%%%%%%%%%%%%%%%%%%%%%%%%%%%%%%%%%%%%%%%%%%%%
%%\subsection{Test Case for TiO2 and Ortho-MAPI}
%%%%%%%%%%%%%%%%%%%%%%%%%%%%%%%%%%%%%%%%%%%%%%%%%%%%%%%%%%%%%%%%%%%%%%%
As a test case of our method, we checked orthorhombic MAPI, whose modes are all non-degenerate, and TiO$_2$, which has some doubly degenerate modes. Our method is able to calculate the mode irreducible representations for these two exactly symmetric structures (Fig. S5), comparable to the Quantum ESPRESSO \texttt{ph.x} output of the mode irreducible representations. We are able to calculate the irreducible representations exactly even without considering the hydrogen atoms in orthorhombic MAPI. This result suggests it is reasonable to try to calculate mode irreducible representations of the tetragonal structure without considering the H atoms.\par

%%-----------------------------
%% irreducible representations for both I4cm and I4mcm structures (give figure)

\begin{figure}[ht!]
\subfloat[I4cm (Pb-I cage only)]{\includegraphics[width =\textwidth ]{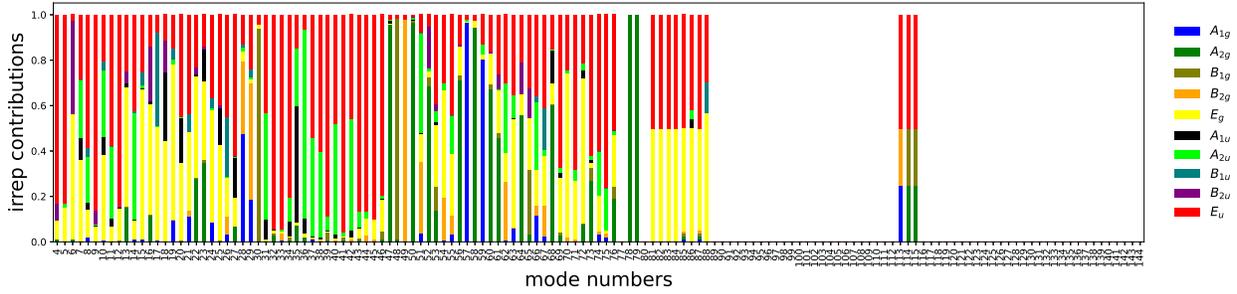}} \\
\subfloat[I4cm complete sructure]{\includegraphics[width =\textwidth]{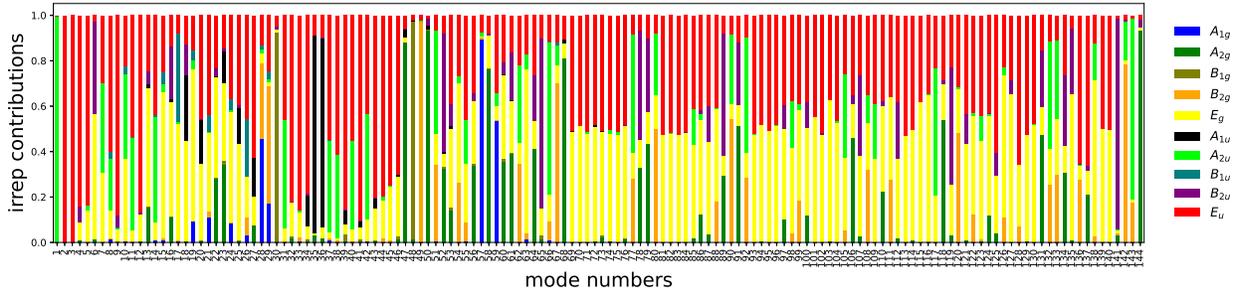}}\\
\subfloat[I4/mcm complete structure]{\includegraphics[width =\textwidth]{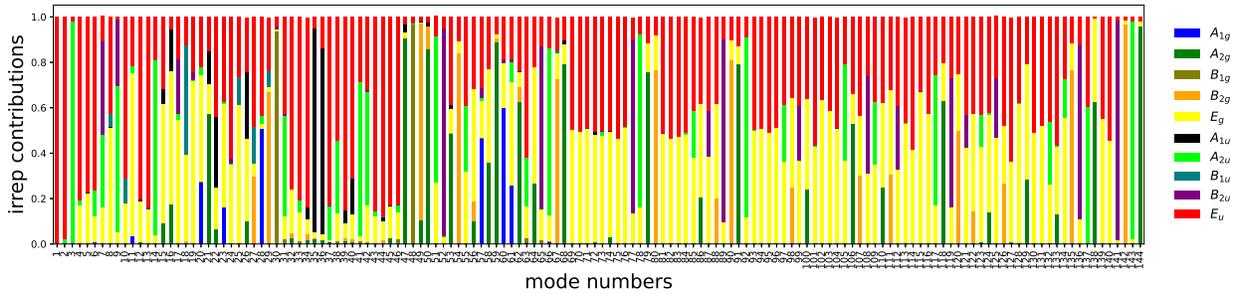}}
\caption{Contributions of different irreducible representations for each mode in (a) Pb-I cage only structure of I4cm symmetry, (b) full I4cm structure, and (c) full I4/mcm structure calculated considering the highest symmetry $D_{4h}$ of the tetragonal structure.}
\label{fig:irrep_contrib}
\end{figure}

%%------------------------------

Now applying the method to tetragonal MAPI: the contribution of irreducible representations for each mode of Pb-I cage is shown in Fig. \ref{fig:irrep_contrib}(a). Because the mid and high frequency modes do not have much Pb-I vibrations it is not enough to get irreducible representations for all the modes of tetragonal MAPI just using only Pb-I cage. It also indicates that the high frequency modes are purely molecular modes. We need to consider the molecular vibrations if we want to calculate the irreducible representations correctly for mid and high frequency modes. We decided to keep the C and N of the MA$^+$ ion with the Pb-I cage and not consider the H atoms which are randomly oriented anyway and hard to track after the rotational symmetry operation on the structure as they are more in number and close to each other in space. Our reason of not considering H is also supported by the idea that, for orthorhombic MAPI we are able to calculate exact irreducible representations for each mode even without considering the H atoms in the structure and we have also checked that the contribution of the H atoms in each mode eigenvectors for both orthorhombic and tetragonal structure looks similar and the H mainly affects the high-frequency modes (Fig. S3). We followed the same process as we mentioned earlier for Pb-I cage and calculated the contributions of the irreducible representations for phonon modes of both I4cm and I4/mcm tetragonal structures. The result is given in Fig. \ref{fig:irrep_contrib}(b,c). We can see that for low-frequency modes, both Pb-I cage-only calculation and the entire structure (except H) give similar result, for mid-frequency region the molecular modes change the irreducible representations that are coming from Pb-I only. It can be also seen that some modes obey the symmetry better than the others.\par
 
 To assess the degree to which each vibrational mode obeys symmetry, we construct a quantity $T \left(C_k \right)$ that is equal to the number of modes $N = 144$ for a perfectly obeyed symmetry operation $C_k$. In the absence of degeneracy, all characters are $\pm 1$, and so the sum of the squared characters $\chi_i$ of modes for any class $C_k$ should be equal to the total number of modes, $T \left(C_k \right) = \sum_{i=1}^N \chi_i^2\left({C_k}\right) = N$. However, degenerate modes should be treated together, as their individual characters are arbitrary and only the sum of their characters is meaningful. For doubly degenerate modes (no triple degeneracies occur in $D_{4h}$), this sum can be $-2$, 0, or 2. In this case, rather than $\chi_i^2 + \chi_{i+1}^2$ in the sum, we use $2 (|\chi_i + \chi_{i+1}| - 1)^2$, which gives a contribution of 2 for the two modes together for anhy of these 3 possible ideal values, and preserves the idea of a total sum of $N$. How do we identify degenerate modes in the presence of approximate symmetry? Almost all the modes have some contributions from the doubly degenerate $E_g$ and $E_u$ representations of $D_{4h}$. We consider a mode degenerate if the sum of the contributions from $E_g$ and $E_u$ is greater than 80\%, and there is a pair of consecutive modes close in frequency. The results for $T \left( C_k \right)$ are given in Fig. S4. We can see that some of the symmetry operations such as $C_4, C^\prime_2, S_4, \sigma_v$, and $\sigma_d$ have values close to 144, while others are as low as half this. We see that $T \left( C_k \right)$ is the same for each member of a class, as expected.\par 

To consider whether the vibrational modes of the ostensibly I4/mcm structure are best described by I4/mcm symmetry or some other subgroup, we have assessed which symmetry operations are obeyed by the modes. For example, we can see that $\sigma_d$ is obeyed better than rest of the operations. So subgroup $C_s$ clearly applies well. To check more rigorously, we have calculated the contribution of irreducible representations of vibrational modes based on each subgroup, and ranked each subgroup based on a value ($RG$) as given in Eq. \ref{eq:RG}. 

\begin{equation}\label{eq:RG}
    RG=\sum^{N_{\rm modes}}_\nu \sum^{N_{\rm irreps}}_i \mu^2_{\nu,i}
\end{equation}
Here $\mu_{\nu,i}$ is the contribution of the irreducible representation $i$ for mode $\nu$, and the squaring is analogous to the inverse participation ratio. The sum should be equal to or less than the total number of modes, which is 144 in our case but will be less as our structure is not properly symmetric. This is because for a perfect irreducible representation of a mode, the maximum value of $\mu_{\nu,i}$ can be 1.

\begin{figure}[ht!]
    \centering
    \includegraphics[width=\textwidth]{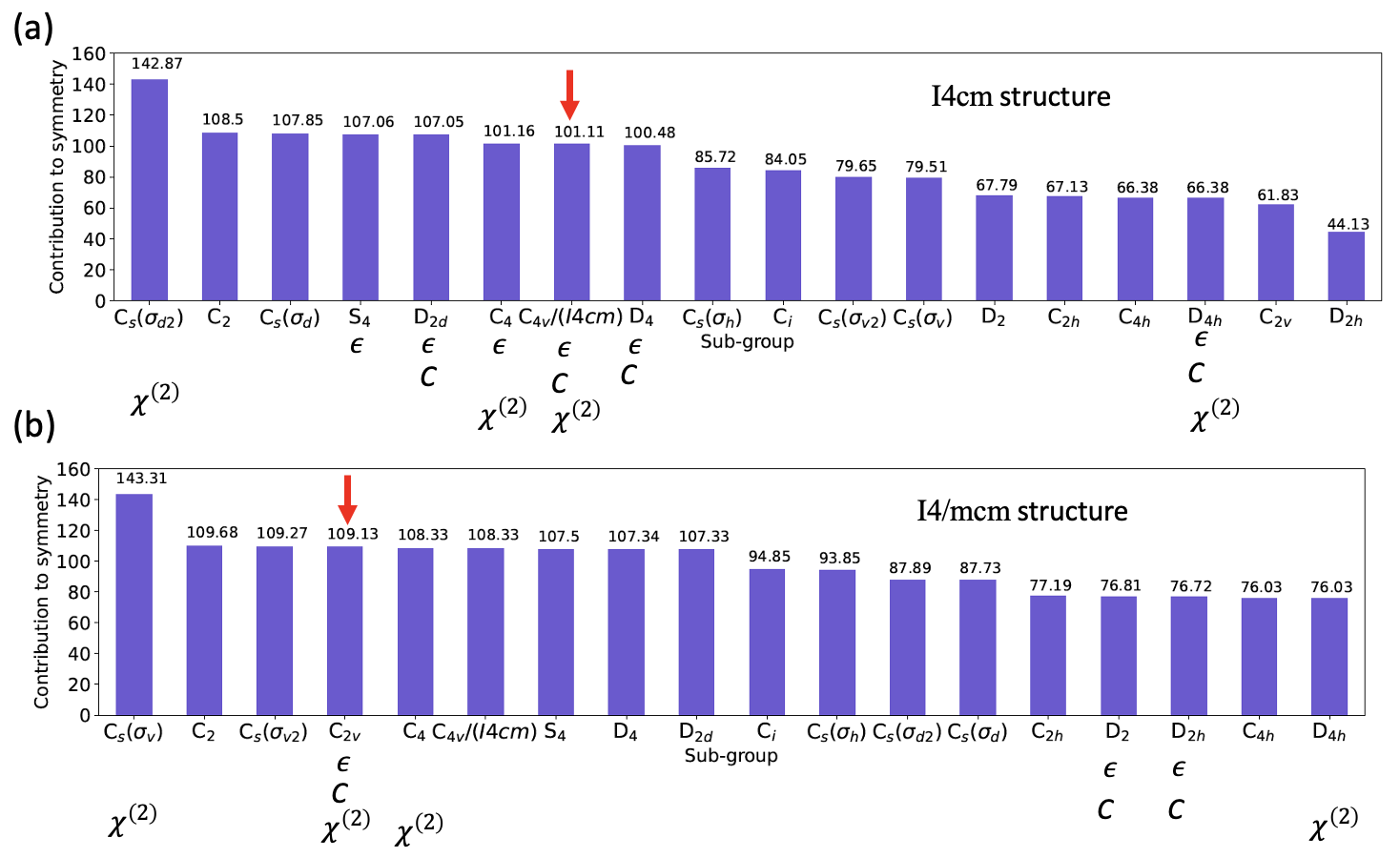}
    \caption{Ranking of subgroups of $D_{4h}$ in declining order of satisfaction of symmetry by the vibrational modes of the two MAPI structures. The subgroups are also annotated with a tensor ($\epsilon$, $C$, and $\chi^{(2)}$) if the tensor's symmetry properties are consistent with that subgroup.}
    \label{Fig:sub_group_rank}
\end{figure}

The plot for the rank of each subgroup is given in Fig. \ref {Fig:sub_group_rank}. Based on the symmetry of the stiffness tensor ($C$), dielectric tensor tensor ($\epsilon$), electro-optic tensor ($\chi^{(2)}$), and calculated rank of the subgroup we can see that C$_{4v}$ is the best symmetry point group for I4cm and C$_{2v}$ for I4/mcm structure. 

Given the limitations of the method above for analyzing symmetry of vibrational modes, we also investigated the symmetry directly of atom-resolved tensors of the system, in particular the Born effective charge tensors ($Z_{\alpha ij}$),  atomic Raman tensors ($R_{ijk\alpha}$) and dynamical matrix ($D^s_{i\alpha j\beta}$). Here $\alpha, \beta$ are the atom indices (including only Pb, I, C, and N atoms) and $i,j$ represent the Cartesian $x$, $y$, and $z$ directions. These tensors are the source of IR and Raman spectroscopy, and in the case of $Z$ and $R$, provide a mixed structural/electronic property. 
Each tensor was calculated using density functional perturbation theory in Quantum ESPRESSO. We can quantify deviations from symmetry by transforming the tensor under a symmetry operation and calculating the deviation from the original. If the structure obeys the symmetry perfectly, then deviations from symmetry $\Delta_Z$ (Eq. \ref {eq:BC}), $\Delta_R$ (Eq. \ref {eq:RT}), and $\Delta_D$ (Eq. \ref{eq:DM}) should be zero for each symmetry operation.
\begin{equation}\label{eq:BC}
    \Delta_Z = \sqrt{\sum_{\alpha ij} \left| Z_{\alpha ij} - \sum_{\alpha' i'j'}K_{ii'} K_{jj'} M_{\alpha\alpha'} Z_{\alpha' i'j'} \right|^2}
\end{equation}
\begin{equation}\label{eq:RT}
   \Delta_R = \sqrt{\sum_{ijk\alpha} \left| R_{ijk\alpha}-\sum_{i'j'k'\alpha'}K_{ii'} K_{jj'}K_{kk'}M_{\alpha \alpha'}R_{\alpha' i'j'k'} \right|^2}
\end{equation}
\begin{equation}\label{eq:DM}
    \Delta_D = \sqrt{\sum_{i j\alpha \beta } \left|D^s_{i\alpha j\beta}-\sum_{i'j'\alpha'\beta'}K_{ii'} M_{\alpha \alpha'}K_{jj'}M_{\beta \beta'}D^s_{\alpha' i' j'\beta'}\right|^2}
\end{equation}
%$M_{ii'}$ represents the rotational matrix for a symmetry operation.
Calculated deviation from symmetry for both I4cm and I4/mcm structures are shown in Fig. \ref {fig:devsym}, giving quite similar values in the two cases. $\sigma_d$ for I4cm and $\sigma_v$ for I4/mcm are obeyed almost perfectly. Here we can see that the symmetries that belong to the same class (e.g. $C_4(1)$ and $C_4(2)$) give the same value for the deviation of symmetry, which is because those operations are related by the highly satisfied symmetries $\sigma_d$ for I4cm and $\sigma_v$ for I4/mcm. To assess the significance of the other deviations, we can compare to the Frobenius norm of each tensor (Eqs. \ref{eq:BCF}, \ref{eq:RTF}, \ref{eq:DMF}).
\begin{equation}\label{eq:BCF}
    ||Z|| = \sqrt{\sum_{\alpha ij} \left| Z_{\alpha ij} \right|^2}
\end{equation}
\begin{equation}\label{eq:RTF}
   ||R|| = \sqrt{\sum_{ijk\alpha} \left| R_{ijk\alpha} \right|^2}
\end{equation}
\begin{equation}\label{eq:DMF}
    ||D|| = \sqrt{\sum_{i j\alpha \beta } \left|D^s_{i\alpha j\beta}\right|^2}
\end{equation} 
We find that $||Z|| = 21.83\ {\rm a.u.}^{-1}$ for I4cm and $21.91\ {\rm a.u.}^{-1}$ for I4/mcm, 
$||R|| = 0.9956\ {\rm a.u.}^{-1}$ for I4cm and $1.024\ {\rm a.u.}^{-1}$ for I4/mcm, and $||D|| = 5.607\ {\rm Ryd}^2$ for I4cm and $5.603\ {\rm Ryd}^2$ for I4/mcm. Deviations are small compared to the Frobenius norm for $Z$ and $D$, indicating all symmetry operations are approximately valid, but on the same order as the Frobenius norm for $R$, indicating poor satisfaction of symmetries. As a result, we can expect approximate symmetries in tetragonal MAPI to be useful for analysis of IR spectroscopy but not very useful for analysis of Raman spectroscopy. The close satisfaction of $\sigma$ symmetries indicates that $C_s$ is an appropriate point group in both cases, but the similar values of the deviation for other operations does not help to distinguish further what higher symmetry may be appropriate, and so our picture of the approximate symmetry remains based on the vibrational modes and global tensors.\par

%%%%%%%%%%%%%%%%%%%%%%%%%%%%%%%%%%%%%%%%%%%%%%%%%%%%%%%%%%%%%%%%%%%%%%%
%% Conclusion
%%%%%%%%%%%%%%%%%%%%%%%%%%%%%%%%%%%%%%%%%%%%%%%%%%%%%%%%%%%%%%%%%%%%%%%
We have analyzed hidden symmetry in theoretical structures of tetragonal MAPI, via a group theory analysis of vibrational modes and by rotation of response tensors, to quantify approximate symmetries that can be used to understand its spectroscopy and other properties. Theoretical calculations have proposed predominant structures referred to as quasi-I4cm or quasi-I4/mcm, but neither possesses any exact symmetry in its atomic coordinates. Nevertheless, by looking at symmetry in various perspectives and considering subgroups of the full tetragonal symmetry ($D_{4h}$ point group), we find that the quasi-I4cm structure can indeed be best described by approximate I4cm ($C_{4v}$ point group) symmetry, whereas the quasi-I4/mcm structure is best described not by I4/mcm but by the lower symmetry of the $C_{2v}$ subgroup. Our methodology allows us to quantify the approximate symmetry of different vibrational modes, by analysis into irreducible representations, and to quantify the degree to which each symmetry operation is satisfied by the modes. We also assessed the symmetry of global response tensors (dielectric, elastic, and electro-optic) and atom-resolved response tensors (Born effective charge, Raman, and dynamical matrix) to develop a combined picture of the usable symmetries in this material. We exclude the H atoms from the analysis due to rapid cation rotations except at very low temperature. Our methodology can be useful to rigorously quantify approximate symmetry in a material, for example doped structures, polycrystalline materials or even amorphous materials and be helpful to understand spectroscopy. Such an approach is particularly important for novel soft semiconductors such as low-dimensional hybrid perovskites\cite{mcclintock2022surface} and other organic metal halide hybrid materials,\cite{lee2022bulk} which typically feature symmetry-breaking cation rotations except at the lowest temperatures, and yet still have enough symmetry for group theory to be useful in analysis. Our methodology can be used for a variety of properties such analysis of strain effects on Raman spectra\cite{talit}, in which we previously established approximate isotropic symmetry in amorphous Si\cite{Strubbe}, or for excited-state forces and exciton-phonon couplings. 
%Our detailed analysis of each symmetry operations that belongs to the highest tetragonal symmetry ($D_{4h}$) reveals that there is not much difference between these two structures with respect to centrosymmetry. After considering all possible lower symmetry subgroups of I4/mcm, we found that $C_{4v}$ or I4cm is the symmetry that best describes the structure and can be used to compare with the experimental structure.
\par
%%%%%%%%%%%%%%%%%%%%%%%%%%%%%%%%%%%%%%%%%%%%%%%%%%%%%%%%%%%%%%%%%%%%%%%
%% Acknowledgement
%%%%%%%%%%%%%%%%%%%%%%%%%%%%%%%%%%%%%%%%%%%%%%%%%%%%%%%%%%%%%%%%%%%%%%%
\begin{acknowledgement}
This material is based upon work supported by the Air Force Office of Scientific Research under award number FA9550-19-1-0236. This work used computational resources from the Multi-Environment Computer for Exploration and Discovery (MERCED) cluster at UC Merced, funded by National Science Foundation Grant No. ACI-1429783, and the National Energy Research Scientific Computing Center (NERSC), a U.S. Department of Energy Office of Science User Facility operated under Contract No. DE-AC02-05CH11231.
\end{acknowledgement}
%%%%%%%%%%%%%%%%%%%%%%%%%%%%%%%%%%%%%%%%%%%%%%%%%%%%%%%%%%%%%%%%%%%%%%%
%% Bibliography
%%%%%%%%%%%%%%%%%%%%%%%%%%%%%%%%%%%%%%%%%%%%%%%%%%%%%%%%%%%%%%%%%%%%%%%
\bibliography{hidden_symm.bib}

\end{document}